\documentclass[aps,reprint,superscriptaddress]{revtex4-1}
\usepackage{graphicx,amsmath,amsfonts,amssymb,esint}

\begin{document}
\title{Slow-wave based magnonic diode}
\author{Mat\'{\i}as Grassi}
\affiliation{Universit\'e de Strasbourg, CNRS, Institut de Physique et Chimie des Mat\'eriaux de Strasbourg, UMR 7504, France}
\author{Moritz Geilen}
\affiliation{Fachbereich Physik and Landesforschungszentrum OPTIMAS, Technische Universit\"{a}t Kaiserslautern, Germany}
\author{Damien Louis}
\affiliation{Universit\'e de Strasbourg, CNRS, Institut de Physique et Chimie des Mat\'eriaux de Strasbourg, UMR 7504, France}
\author{Morteza Mohseni}
\affiliation{Fachbereich Physik and Landesforschungszentrum OPTIMAS, Technische Universit\"{a}t Kaiserslautern, Germany}
\author{Thomas Br\"{a}cher}
\affiliation{Fachbereich Physik and Landesforschungszentrum OPTIMAS, Technische Universit\"{a}t Kaiserslautern, Germany}
\author{Michel Hehn}
\affiliation{Institut Jean Lamour, Universit\'e de Lorraine, UMR 7198 CNRS, Nancy, France}
\author{Daniel Stoeffler}
\affiliation{Universit\'e de Strasbourg, CNRS, Institut de Physique et Chimie des Mat\'eriaux de Strasbourg, UMR 7504, France}
\author{Matthieu Bailleul}
\affiliation{Universit\'e de Strasbourg, CNRS, Institut de Physique et Chimie des Mat\'eriaux de Strasbourg, UMR 7504, France}
\author{Philipp Pirro}
\affiliation{Fachbereich Physik and Landesforschungszentrum OPTIMAS, Technische Universit\"{a}t Kaiserslautern, Germany}
\author{Yves Henry}
\affiliation{Universit\'e de Strasbourg, CNRS, Institut de Physique et Chimie des Mat\'eriaux de Strasbourg, UMR 7504, France}
\date{\today}

\begin{abstract}
Spin waves, the collective excitations of the magnetic order parameter, and magnons, the associated quasiparticles, are envisioned as possible data carriers in future wave-based computing architectures. On the road towards spin-wave computing, the development of a diode-like device capable of transmitting spin waves in only one direction, thus allowing controlled signal routing, is an essential step. Here, we report on the design and experimental realization of a microstructured magnonic diode in a ferromagnetic bilayer system. Effective unidirectional propagation of spin waves is achieved by taking advantage of nonreciprocities produced by dynamic dipolar interactions in transversally magnetized media, which lack symmetry about their horizontal midplane. More specifically, dipolar-induced nonreciprocities are used to engineer the spin-wave dispersion relation of the bilayer system so that the group velocity is reduced to very low values for one direction of propagation, and not for the other, thus producing unidirectional slow spin waves. Brillouin light scattering and propagating spin-wave spectroscopy are used to demonstrate the diode-like behavior of the device, the composition of which was previously optimized through micromagnetic simulations.
\end{abstract}

\maketitle

\section{Introduction}\label{Sec_Intro}
The isolator, or wave diode, is an essential building block in wave computing architectures, which is necessary to mitigate unwanted reflections and prevent signal backflow\,\cite{JPEF13}. To act as a diode, that is, to transmit waves in only one direction, and, more generally, to exhibit nonreciprocity, a wave device operated in the linear regime must break time-reversal ($T$) symmetry\,\cite{LL60}. While breaking of $T$ symmetry in photonic and electronic systems requires complex time-dependent externally driven modulations, like propagating refractive index perturbation\,\cite{LYFL12} or staggered commutation\,\cite{RK16}, it happens readily in gyrotropic systems, such as magnetically biased two-dimensional electron gases\,\cite{MCPH17} or plasmas\,\cite{GM18}, and it exists intrinsically in ferro- and ferrimagnetic materials, which, in a bulk form, constitute the heart of standard microwave isolators\,\cite{FC65}.

Magnetic materials host spin waves with typical frequencies and wavelengths in the giga-terahertz and nano-micrometer ranges, respectively, which are considered as possible carriers for wave computing or fast information processing in so-called magnonic circuits\,\cite{KW11}. In order to avoid power losses upon signal conversion to the electrical domain, an all-magnon approach\,\cite{C19}, where full data processing would be performed within the spin-wave domain, would be highly desirable. In recent years, important steps have been taken in this direction, which include the experimental demonstration of a magnonic majority gate\,\cite{FKBS17} and the design of a magnonic half-adder\,\cite{WVBP19}. Designs of magnonic diodes\,\cite{LYWX15} and unidirectional spin-wave emitters\,\cite{BBGP17} relying on the chiral Dzyaloshinskii-Moriya interaction for producing nonreciprocity have been proposed theoretically, but have not been realized yet. Very recently, however, a unidirectional spin-wave emitter whose working principle is based on chiral magneto-dipolar interactions has been demonstrated experimentally\,\cite{CYLL19}.

In the present paper, we study both theoretically and experimentally a new kind of magnonic diode based on the concept of slow spin waves and relying on dipolar-induced nonreciprocity. Key to the operation of the device is the Damon-Eshbach (DE) or magnetostatic surface wave (MSSW) configuration, where the equilibrium magnetization of a thin film lies in-plane and at right angle from the direction of spin-wave propagation\,\cite{DE61}. This configuration is specific in that a dipolar coupling exists between the two (in-plane and out-of-plane) components of the dynamic magnetization\,\cite{K13}, which happens to be nonreciprocal. In the presence of top/bottom magnetic asymmetry, this coupling does not average out to zero and translates into counterpropagating surface waves with a given wave number having different frequencies\,\cite{H90,KRVZ07,HBKL14,GHHK16,GASG19}.

Here, we engineer the spin-wave dispersion of a thin film system consisting of two exchange-coupled ferromagnetic layers with different magnetic parameters so as to bring the group velocity of spin waves travelling in a particular direction to very low values, while keeping those of counterpropagating waves large enough to ensure that they travel over long distances. We use Brillouin light scattering and propagating spin-wave spectroscopy\,\cite{BOF03} to demonstrate that these nonreciprocal slow spin waves allows one to build an efficient spin-wave diode out of a Co$_{40}$Fe$_{40}$B$_{20}$/Ni$_{80}$Fe$_{20}$ bilayer waveguide. Two types of numerical simulations are used to support our experimental work. An in-house developed finite difference method, which performs a plane spin-wave normal mode analysis\,\cite{HGB16}, is employed to compute dispersion relations and to optimize the composition of the investigated bilayer system prior to its fabrication. The MuMax3 program\,\cite{VLDH14}, on the other hand, is used to produce magnetization maps, mostly for illustration purposes. Importantly, an analytical model is first presented for explaining in some depth the functioning of the proposed magnonic diode and shed light on the specific chiral dipolar couplings on which it relies.

\section{Design principle and theory}\label{Sec_Design&Theory}
\subsection{Design principle}\label{Sec_Design}
\begin{figure}[t]
\includegraphics[width=8.5cm,trim=55 165 65 190,clip]{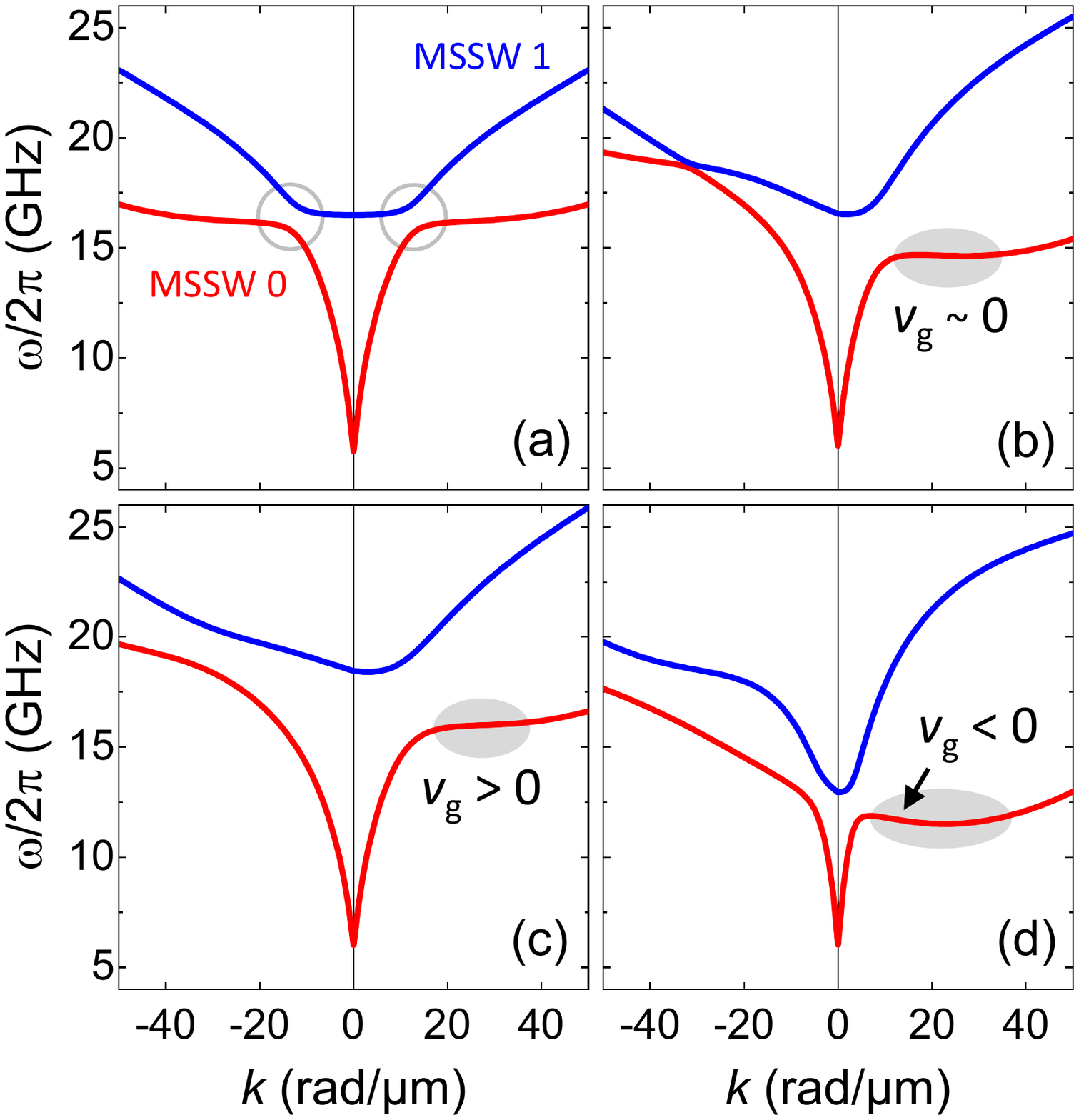}
\caption{(a) Computed dispersion relations for the two lowest frequency MSSW modes in a 38~nm thick single layer with an exchange constant of 15~pJ/m and a saturation magnetization of 1060~kA/m submitted to a transverse applied magnetic field $\mu_0H_0=30$~mT. The gray circles indicate mode anticrossings. (b-d) Same as in (a) for CoFeB/Py bilayers with (b) $l_{\text{CoFeB}}\!=\!l_{\text{Py}}\!= \!19\,\text{nm}$, (c) $l_{\text{CoFeB}}\!=\!l_{\text{Py}}\!=\!17\,\text{nm}$, and (d) $l_{\text{CoFeB}}\!=\!l_{\text{Py}}\!=\!25\,\text{nm}$. The gray ellipses highlight regions where the group velocity of MSSW mode 0 is small.}\label{Fig_Principle}
\end{figure}

The bilayer system that we consider has a total thickness $l$ of several tens of nanometers. For such a thickness, the two lowest frequency MSSW branches are close to each other and hybridize, as revealed by the appearance of mode anticrossings in their dispersion [Fig.~\ref{Fig_Principle}(a)]. This hybridization plays a crucial role in obtaining plateaus in the dispersion, which are synonymous of spin waves with vanishing group velocity. For a magnetically symmetric film, the hybridization effect is similar for both positive and negative values of the in-plane wave vector $k$ [Fig.~\ref{Fig_Principle}(a)]. In a bilayer film made of two magnetic materials with different saturation magnetization $M_{\text{S}}$, on the contrary, mode coupling is nonreciprocal [Fig.~\ref{Fig_Principle}(b)]: For one sign of $k$ (positive here), modes repel each other (much) more strongly than for the other sign\,\cite{KK90}. This asymmetry leads quite systematically to the existence of a rather wide $k$-range where the first MSSW mode is only weakly dispersive and group velocity is low [gray shaded regions in Fig.~\ref{Fig_Principle}(b-d)], while it remains large for counterpropagating waves with the same frequency. For large thicknesses, mode repulsion can even lead to surface waves exhibiting an unusual backward character (phase velocity $v_p\!=\!\omega/k$ and group velocity $v_g\!=\!\partial w/\partial k$ with opposite signs), see arrow in Fig.~\ref{Fig_Principle}(d). We note that a similar backward inflexion associated with the existence of a nonreciprocal magnetostatic interface mode\,\cite{W70,CM82} was also predicted for magnetic bilayers in the hypothetical pure dipolar regime\,\cite{RWJ87}.

The nonreciprocal flattening of the dispersion relation of MSSW modes in bilayers is quite generic. A priori, it can be obtained using any two magnetic materials with sufficiently different $M_{\text{S}}$ values\,\cite{Rem1}. Once these are chosen, designing an optimal diode device amounts to adjusting the composition of the bilayer (total thickness $l$ and/or individual thicknesses of the two layers $l_j$, with $j\!=\!1,2$) so as to obtain a frequency plateau as flat as possible [Fig.~\ref{Fig_Principle}(b-d)]. Numerical simulations show that this can usually be achieved in many ways (not shown). With the ferromagnetic alloys chosen for the present study, namely Co$_{40}$Fe$_{40}$B$_{20}$ ($M_{\text{S,CoFeB}}\!=\!1270$~kA/m, $A_{\text{CoFeB}}\!=\!17$~pJ/m) and Permalloy ($M_{\text{S,Py}}\!=\!845$~kA/m, $A_{\text{Py}}\!=\!12.8$~pJ/m), $l_{\text{CoFeB}}\!=\!l_{\text{Py}}\!= \!19\,\text{nm}$ is a good choice [Fig.~\ref{Fig_Principle}(b)]. In the forthcoming experimental section (Sec.~\ref{Sec_Exp_Results}), we shall use a different composition of the bilayer, namely $\{l_{\text{CoFeB}}\!=\!20\,\text{nm}, l_{\text{Py}}\!=\!26\,\text{nm}\}$, which yields a slightly broader plateau.

\subsection{Theoretical model}\label{Sec_Theory}
\begin{figure}
\includegraphics[width=6cm,trim=155 85 140 580,clip]{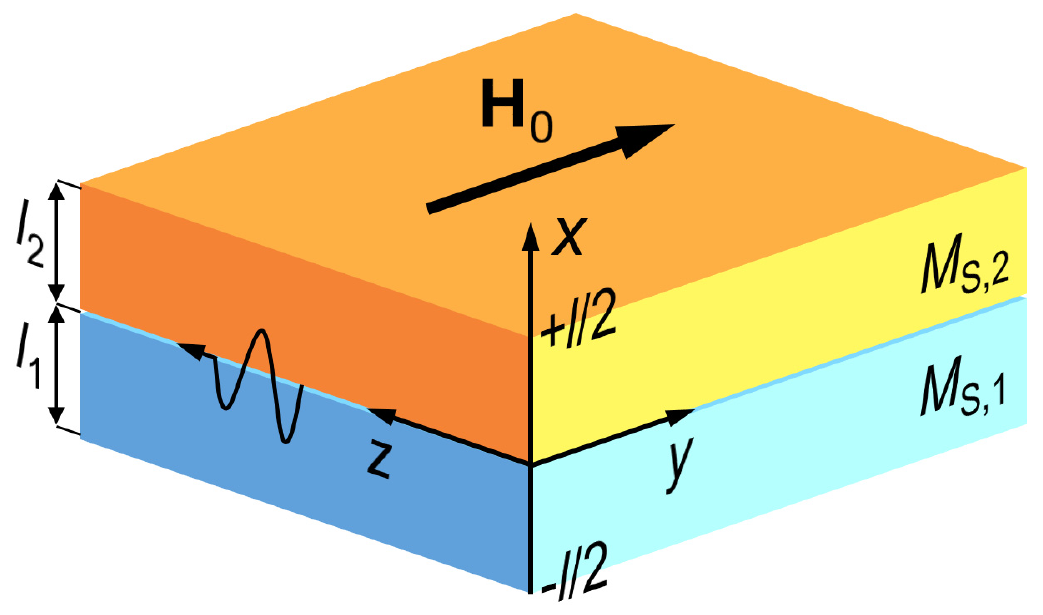}
\caption{Schematic representation of the bilayer system considered in the analytical model of Sec.~\ref{Sec_Theory}.}\label{Fig_Bilayer_System}
\end{figure}

To describe the peculiar magnetization dynamics in a bilayer film and to account for the occurrence of nonreciprocal mode hybridization, we may resort to the theory of dipole-exchange spin waves\,\cite{KS86} and use a similar analytical approach as previously developed to account for frequency nonreciprocities of surface waves induced by asymmetric surface anisotropies\,\cite{GHHK16}. Here, for the sake of simplicity, we make two strong assumptions: i) First, we take $l_1\!=\!l_2$, which allows us to decompose the profile of the saturation magnetization through the film thickness $M_{\text{S}}(x)$ into the sum of a mean $\langle M_{\text{S}} \rangle = (M_{\text{S,1}} + M_{\text{S,2}})/2$ and an antisymmetric deviation $\langle M_{\text{S}}\rangle \,a(x)$, where $a(x)\!=\!\beta\,\text{sgn}(x)$ and $\beta\!=\!\frac{M_{\text{S,1}}-M_{\text{S,2}}}{M_{\text{S,1}} + M_{\text{S,2}}}$ denotes the contrast of saturation magnetization between the two materials; ii) Second, we assume that the ratio of the exchange constant to the saturation magnetization takes a constant value $A_1/M_{\text{S,1}}\!=\!A_2/M_{\text{S,2}}\!=\!\langle A\rangle/\langle M_{\text{S}}\rangle$ throughout the film. In these conditions, the linearized Landau-Lifshitz equation for plane spin waves of the form $\mathbf{m}(x,z,t) = M_{\text{S}}(x)\,\mathbf{n}(x,z,t)=M_{\text{S}}(x)\,\boldsymbol{\eta}(x) \,e^{i(\omega t-kz)} $ can be written as
\begin{align}\label{Eq_LL}
i\omega\boldsymbol{\eta}(x)\!=
&-\gamma\mu_0H_0\,\boldsymbol{\eta}(x)\times\hat{\mathbf{y}}\nonumber\\
&+\gamma\frac{2\langle A\rangle}{\langle M_{\text{S}}\rangle}\left(\frac{\partial^2}{\partial x^2}-k^2\right)\,\boldsymbol{\eta}(x)\times\hat{\mathbf{y}}\\
&+\gamma\mu_0\langle M_{\text{S}}\rangle\!\int_{\!-l/2}^{l/2}\! \bar{\bar{G}}_k(x\!-\!x')[1\!+\!a(x')](\boldsymbol{\eta}(x')\!\times\hat{\mathbf{y}}) \,dx'\nonumber
\end{align}
where $\gamma$ is the gyromagnetic ratio\,\cite{Rem2}, $\mu_0$ is the permittivity of vacuum, $\hat{\mathbf{y}}$ is a unit vector along the applied magnetic field $\mathbf{H}_0$ [Fig.~\ref{Fig_Bilayer_System}], and $\bar{\bar{G}}_k$ is the magnetostatic Green's function. Using as a vector basis the $x$ and $z$ components of the first two unpinned exchange modes with homogeneous ($n\!=\!0$) and fully antisymmetric ($n\!=\!1$) profiles of the magnetization precession angle across the film thickness\,\cite{Rem3}, that is, the four vector set $\{S_0\,\hat{\mathbf{x}},S_0\,\hat{\mathbf{z}},S_1\,\hat{\mathbf{x}}, S_1\,\hat{\mathbf{z}}\}$ with $S_0(x)\!=\!\frac{1}{\sqrt{l}}$ and $S_1(x)\!=\!\sqrt{\frac{2}{l}}\,\sin(\frac{\pi x}{l})$, Eq.~\ref{Eq_LL} can be rewritten in the form of an eigenvalue equation $i\Omega\,\bar{\eta}\!=\!\bar{\bar{C}}\,\bar{\eta}$, where $\Omega\!=\!\omega/(\gamma\mu_0\langle M_{\text{S}}\rangle)$ is a dimensionless frequency, $\bar{\eta}\!=\!(\eta_{0,x},\eta_{0,z},\eta_{1,x},\eta_{1,z}) ^{\text{T}}$, and $\bar{\bar{C}}$ is a dynamic matrix\,\cite{GHHK16}.

As an opportunity to introduce useful notations, we first discuss the case of a single layer with exchange constant $\langle A\rangle$ and saturation magnetization $\langle M_{\text{S}}\rangle$ ($\beta\!=\!0$). For such "unperturbed" system, the $\bar{\bar{C}}$ matrix writes
\begin{equation}\label{Eq_Dynamic_Matrix_SL}
\bar{\bar{C}}=
\begin{pmatrix}
\bar{\bar{C}}_{00}&\bar{\bar{C}}_{01}\\\bar{\bar{C}}_{10}&\bar{\bar{C}}_{11}
\end{pmatrix}
=
\begin{pmatrix}
0&\Omega_{0,z}&-iQ&0\\
-\Omega_{0,x}&0&0&iQ\\
iQ&0&0&\Omega_{1,z}\\
0&-iQ&-\Omega_{1,x}&0\\
\end{pmatrix},
\end{equation}
with
\begin{eqnarray}\label{Eq_Omega}
&\Omega_{0,x}&=1-P_{00}+h+\Lambda^{2}k^{2}, \\
&\Omega_{0,z}&=P_{00}+h+\Lambda^{2}k^{2}, \nonumber\\
&\Omega_{1,x}&=1-P_{11}+h+\Lambda^{2}k^{2}+\frac{\Lambda^{2}\pi^{2}}{l^{2}}, \nonumber\\
&\Omega_{1,z}&=P_{11}+h+\Lambda^{2}k^{2}+\frac{\Lambda^{2}\pi^{2}}{l^{2}}, \nonumber
\end{eqnarray}
where $1\!-\!P_{00}\!=\!\frac{1-e^{-|k|l}}{|k|l}$, $P_{00}$, $1\!-\!P_{11}$, and $P_{11}\!=\!\frac{k^2l^2}{\pi^2+k^2l^2} (1-\frac{2\,k^2l^2}{\pi^2+k^2l^2} \frac{1+e^{-|k|l}}{|k|l})$ are $k$-dependent self-demagnetizing factors for the $S_0\hat{\mathbf{x}}$, $S_0\hat{\mathbf{z}}$, $S_1\hat{\mathbf{x}}$, and $S_1\hat{\mathbf{z}}$ basis components, respectively, $h\!=\!H_0/\langle M_{\text{S}}\rangle$ is a dimensionless applied magnetic field and $\Lambda^2\!=\!2\langle A\rangle/(\mu_0\langle M_{\text{S}}\rangle^2)$. The coupling between the uniform and antisymmetric sets of basis functions is described through the off-diagonal $\bar{\bar{C}}_{01}\!=\!\bar{\bar{C}}_{10}^*$ blocks, which involve the nonreciprocal mutual demagnetizing factor $Q\!=\!\frac{\sqrt{2}kl}{\pi^2+k^2l^2} (1+e^{-|k|l})$. Solving for $\text{det}(\bar{\bar{C}}\!-\! i\Omega\bar{\bar{1}})\!=0$, where $\bar{\bar{1}}$ is the identity matrix, yields the eigenfrequencies of the first two hybrid MSSW modes in a single layer
\begin{align}\label{Eq_Eigenfrequencies_SL}
\Omega_{0,1}^2=&\frac{\Omega_{00}^2\!+\Omega_{11}^2}{2}-Q^2\\
&\mp\frac{1}{2}\sqrt{\left(\Omega_{11}^2\!-\Omega_{00}^2\right)^2 + 4Q^2\!\left[(P_{00}\!-\!P_{11})^2\!-\!\frac{\Lambda^4\pi^4}{l^4}\right]},
\nonumber
\end{align}
where $\Omega_{nn}\!=\!\sqrt{\Omega_{n,x}\Omega_{n,z}}$ ($n\!=\!0,1$) are the dimensionless frequencies of the uncoupled uniform and antisymmetric precession modes, $\bar{\eta}_0\!=\!(\eta_{0,x},\eta_{0,z},0,0)^{\text{T}}$ and $\bar{\eta}_1\!=\!(0,0,\eta_{1,x},\eta_{1,z})^{\text{T}}$, respectively. Noticeably, the frequencies of the hybrid modes, $\Omega_0$ and $\Omega_1$, depend on $Q^2$, meaning that, although $Q$ obeys $Q(-k)\!=\!-Q(k)$, hybridization does not effectively produce nonreciprocity [Fig.\ref{Fig_Principle}(a)]. The basic reason for this is that hybridization implies an action of mode $\bar{\eta}_m$ on mode $\bar{\eta}_n$ (for instance, under the influence of mode $\bar{\eta}_1$, mode $\bar{\eta}_0$ acquires an antisymmetric component) as well as a back-action of mode $\bar{\eta}_n$ on mode $\bar{\eta}_m$, both of which are proportional to the coupling factor $Q$.

With these considerations in mind, we are now ready to consider the case of a magnetic bilayer. When $\beta\!\neq\!0$, the diagonal elements in the $\bar{\bar{C}}_{nn}$ blocks and the off-diagonal elements in the $\bar{\bar{C}}_{n\neq m}$'s are no longer zero and the dynamic matrix becomes (see Appendix)
\begin{equation}\label{Eq_Dynamic_Matrix_BL}
\bar{\bar{C}}(\beta) = \bar{\bar{C}}(0) +
\begin{pmatrix}
-iP'_{00}&0&0&Q'\\
0&iP'_{00}&Q'\!-\!I'&0\\
0&Q'&-iP'_{11}&0\\
Q'\!-\!I'&0&0&iP'_{11}\\
\end{pmatrix},
\end{equation}
where $\bar{\bar{C}}(0)$ is given by Eq.~\ref{Eq_Dynamic_Matrix_SL} and
\begin{eqnarray}\label{Eq_beta}
&P'_{00}&=\frac{1}{kl}\left[\sinh(|k|l)-2\sinh\left(\frac{|k|l}{2}\right) \right]\beta, \\
&P'_{11}&= \frac{2kl}{\left(\pi^2\!+\!k^2l^2\right)^2}\left( \pi\!-\!|k|l\,e^{-|k|l\!/2} \right)^2\beta \nonumber\\
&I'&= \frac{2\sqrt{2}}{\pi}\;\beta, \nonumber\\
&Q'&=\frac{2\sqrt{2}}{\pi^2\!+\!k^2l^2} \left[ \pi\!\left( 1\!-\!e^{-|k|l\!/2} \right)+ k^2l^2\!\left( \frac{1}{\pi}-\frac{1\!-\!e^{-|k|l}}{2|k|l} \right) \right]\beta.
\nonumber
\end{eqnarray}
Not surprisingly, all newly non-zero matrix elements are proportional to the contrast in saturation magnetization $\beta$. The coefficients $P'_{nn}$ describe the additional self-demagnetizing effects produced by the magnetic asymmetry on uncoupled modes $\bar{\eta}_n$ ($n=0,1$). As revealed by the imaginary character of the corresponding matrix elements in the $\bar{\bar{C}}_{nn}$ blocks, these take the form of transverse dipole fields, which oscillate with a phase difference of $\pi/2$ with respect to the dynamic magnetization components creating them. The coefficients $Q'$ and $I'$, on the other hand, account for the additional mutual-demagnetizing effects produced in the presence of magnetic asymmetry. They describe how hybridization between the uniform and fully antisymmetric modes is affected in a bilayer. The coefficient $I'$, in particular, corresponds to the local part of the dipole interaction, that is, the usual perpendicular-to-plane demagnetizing effect (see Appendix). It does not depend on $k$ and plays a similar role as a difference in magnetic anisotropies in the top and bottom parts of the film\,\cite{GHHK16}. Therefore, by analogy, we expect it to be responsible for frequency nonreciprocity.

\begin{figure}[t]
\includegraphics[width=8.3cm,trim=35 60 40 50,clip]{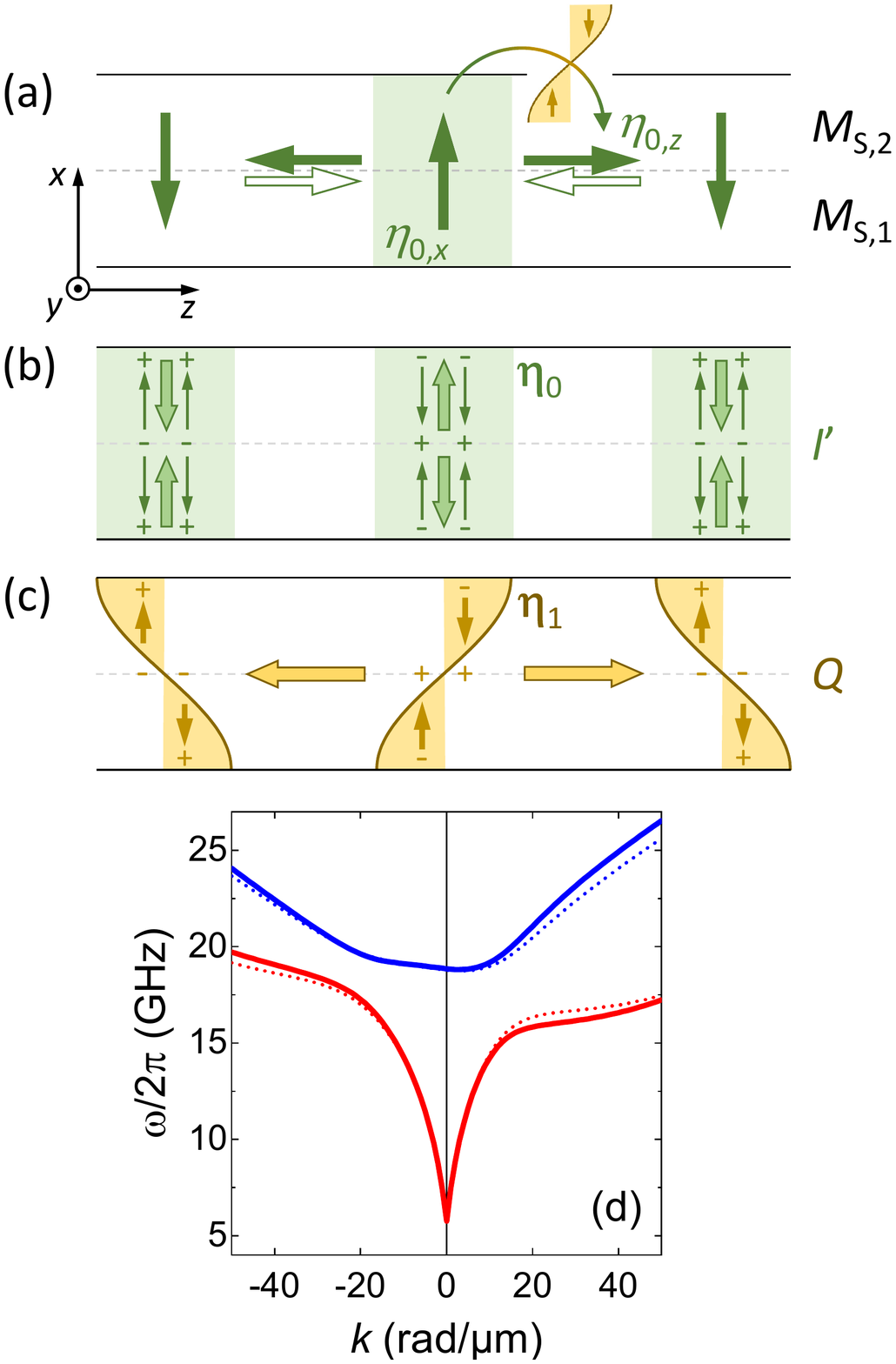}
\caption{(a) Indirect $QI'$ coupling between the in-plane and out-of-plane components of uniform mode $\bar{\eta}_0$, through hybridization with antisymmetric mode $\bar{\eta}_1$. Schematic representation of $\bar{\eta}_0$ for positive $k$ (solid arrows) and negative $k$ (open arrows). (b) Illustration of the $I'$ coupling ($\beta\!>\!0$). For a bilayer, the homogeneous precession angle in $\bar{\eta}_{0}$ translates into an asymmetric distribution of dynamic magnetization. The out-of-plane component of its antisymmetric part (thin green arrows) generates additional magnetic charges and, in turn, an antisymmetric dipole field (thick green arrows), which couples to the out-of-plane component of $\bar{\eta}_{1}$. (c) Illustration of the $Q$ coupling. The out-of-plane component of $\bar{\eta}_{1}$ (thin yellow arrows) produces a symmetric in-plane dipole field (thick yellow arrows), which couples to the in-plane component of $\bar{\eta}_{0}$, in opposite ways depending on the sign of $k$ [compare the relative orientation of the thick horizontal arrows in panels (a) and (c)]. (d) Dispersion relations of the two lowest MSSW modes in a bilayer with $l_1\!+l_2\!=\!34$~nm, $\langle M_{\text{S}}\rangle\!= \!1000$~kA/m, $\langle A\rangle\!=\!15$~pJ/m, and $\beta\!=\!0.1$. Predictions of the analytical model (Eqs.~\ref{Eq_Eigenfrequencies_BL} and \ref{Eq_Delta_Omega}) and results of numerical simulations are shown as solid and dotted lines, respectively.}\label{Fig_Theory}
\end{figure}

\begin{figure*}
\includegraphics[width=14.0cm,trim=35 230 30 200,clip]{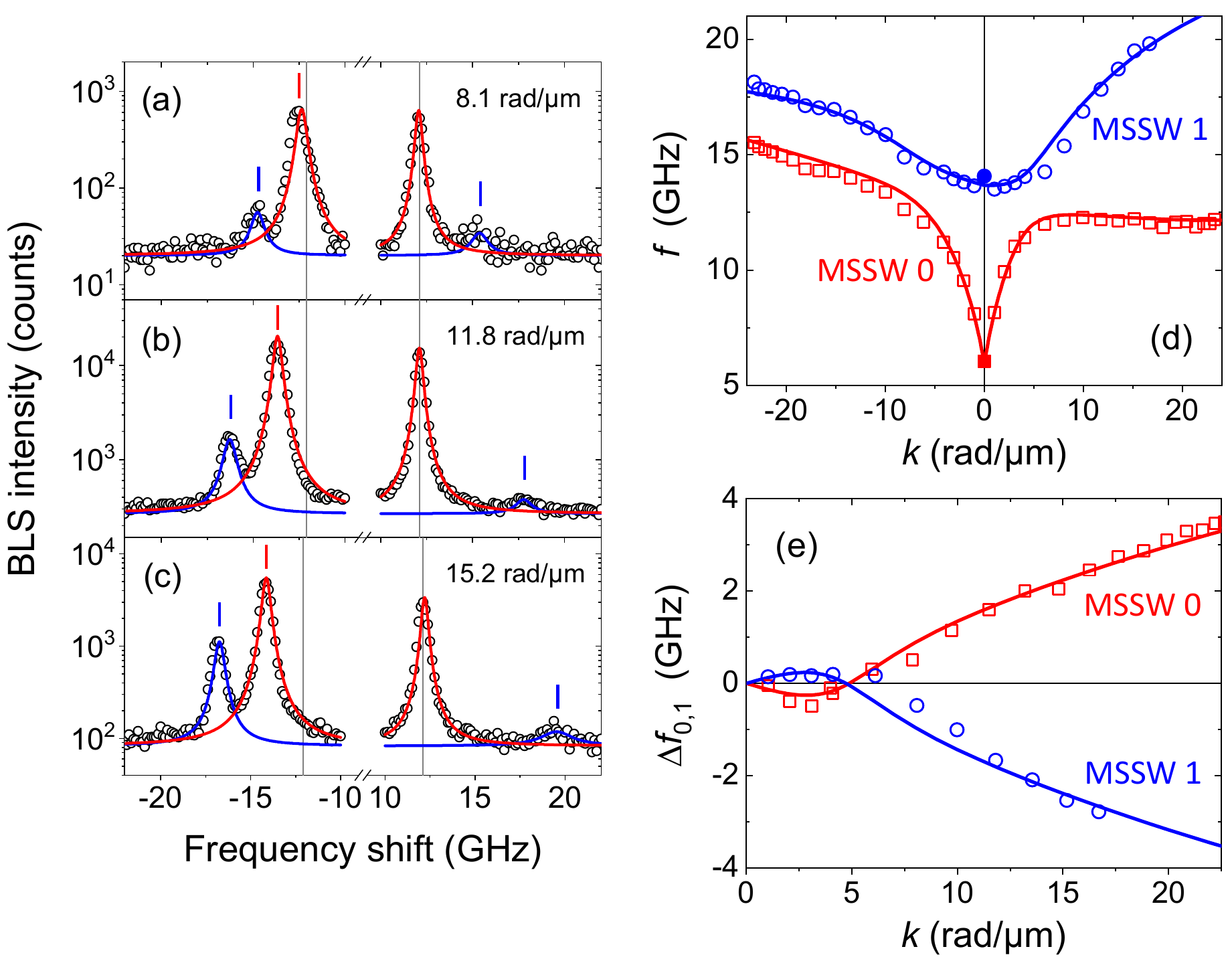}
\caption{(a-c) BLS spectra recorded in the DE configuration ($\mu_0H_0\!=\!+30$~mT) on a Co$_{40}$Fe$_{40}$B$_{20}$(20nm)/Ni$_{80}$Fe$_{20}$(26nm) bilayer film (symbols), for three different values of the in-plane wave vector: (a) $|k|\!=\!8.1$~rad/$\mu$m, (b) $|k|\!=\!11.8$~rad/$\mu$m, and (c) $|k|\!=\!15.2$~rad/$\mu$m. Lines are guides to the eye. (d) Dispersion relations of the two lowest MSSW modes deduced from BLS data (open symbols) and comparison with predictions from numerical simulations (lines). Solid symbols correspond to data from a complementary ferromagnetic resonance experiment. (e) Frequency nonreciprocities $\Delta f_n(k)=f_n(-|k|)-f_n(+|k|)$ of MSSW modes 0 and 1. As in (d), symbols and lines correspond to experimental and numerical data, respectively.}\label{Fig_BLS}
\end{figure*}

Solving for $\text{det}[\bar{\bar{C}}(\beta)\!-\! i\Omega\bar{\bar{1}}]\!=0$ and keeping only terms up to first order in $\beta$ yields the following dispersion relation for the two lowest MSSW modes in a bilayer\,\cite{Rem4}
\begin{eqnarray}\label{Eq_Dispersion_Relation_BL}
&(\Omega_0^2-\Omega^2)(\Omega_1^2-\Omega^2) \\
&-2Q\Omega\left[(Q'\!-\!I')(\Omega_{1,z}-\Omega_{0,z}) + Q'(\Omega_{1,x}-\Omega_{0,x}) \right] = 0. \nonumber
\end{eqnarray}
Treating further $QQ'$ and $QI'$ as small parameters, in a perturbative manner, the eigenfrequencies of these modes can be obtained as
\begin{equation}\label{Eq_Eigenfrequencies_BL}
\Omega_{0,1}(\beta)=\Omega_{0,1}(0)\pm\Delta\Omega(\beta)
\end{equation}
where $\Omega_0(0)$ and $\Omega_1(0)$ are given by Eq.~\ref{Eq_Eigenfrequencies_SL} and
\begin{eqnarray}\label{Eq_Delta_Omega}
\Delta\Omega(\beta)=\frac{Q}{\Omega_1^2(0)\!-\!\Omega_0^2(0)}\,
[(I'\!-Q')&(&\Omega_{1,z}\!-\!\Omega_{0,z})\\
-Q'&(&\Omega_{1,x}\!-\!\Omega_{0,x})] \nonumber.
\end{eqnarray}
The expression of $\Delta\Omega$ involves two products of dipolar matrix elements, $QQ'$ and $QI'$. In both of them, one term ($Q'$ or $I'$) is even in $k$ and the other ($Q$) is odd. Then $\Delta\Omega$ is odd in $k$ and, since $\Omega_0(0)$ and $\Omega_1(0)$ are fully reciprocal, $|2\Delta\Omega|$ is a measure of the frequency nonreciprocities for the two hybrid MSSW modes. Noticeably, the nonreciprocities of these hybrid modes have the same magnitude but opposite signs. In both $QQ'$ and $QI'$ products, also, one term ($Q'$ or $I'$) connects the same components (either $x$ or $z$) for the two uncoupled modes $\bar{\eta}_n$ ($n\!=\!0,1)$, while the other term ($Q$) connects two different components ($x$ and $z$). These peculiar combinations allow for indirect chiral couplings (of dipolar origin) to develop between the two components of each mode $\bar{\eta}_n$, through hybridization with mode $\bar{\eta}_{m \neq n}$. In Fig.~\ref{Fig_Theory}(a-c), we illustrate this crucial point in the simplest case of the $\bar{\eta}_0$ mode and of the $QI'$ product, which plays by far the most important role among the two products, as one may easily verify numerically. Figure~\ref{Fig_Theory}(d) compares predictions of the above analytical model (Eqs.~\ref{Eq_Eigenfrequencies_BL} and \ref{Eq_Delta_Omega}) with results from numerical simulations in the case of a bilayer with $l_1\!+l_2\!=\!34$~nm, $\langle M_{\text{S}}\rangle\!=\!1000$~kA/m, $\langle A\rangle\!=\!15$~pJ/m, and $\beta\!=\!0.1$. A rather good agreement is observed, which confirms the essential role played by the $QQ'$ and $QI'$ chiral couplings identified through the analytical modelling in producing nonreciprocal mode repulsion. We note that these chiral couplings bear some resemblance with those at play in the unidirectional emission of spin waves by nanoscale magnetic transducers\,\cite{CYLL19,AADD12}. Here, however, they are not introduced through transduction (excitation/detection) but they are an intrinsic property of the medium, which supports spin wave propagation.

\section{Experimental results}\label{Sec_Exp_Results}
\subsection{Sample fabrication}\label{Sec_Samples}
The bilayer films used in the present work have been deposited on natively oxidized intrinsic (100)Si substrates by DC magnetron sputtering from material targets with nominal compositions Co$_{40}$Fe$_{40}$B$_{20}$ and Ni$_{80}$Fe$_{20}$. Deposition of the magnetic stack was preceded by that of a 3~nm thick Ta seed layer, for ensuring low layer roughness, and followed by that of a 3~nm thick Au overlayer, for protecting the magnetic alloys against oxydation. For propagating spin-wave spectroscopy and micro-focussed Brillouin light scattering experiments, devices have been fabricated out of these films by means of standard cleanroom processes, involving laser and electron beam lithographies, as well as ion milling\,\cite{HBKL14,VB10}. These consist of $10~\mu$m wide magnonic waveguides covered with a 120~nm thick insulating layer of silicon oxide and two pairs of emitting/receiving microwave antennas, made from a Ti(10nm)/Al(90nm) stack, placed above (see Sec.~\ref{Sec_Diode}).

\subsection{Nonreciprocal dispersion relations}\label{Sec_BLS}
The dispersion relations of thermally excited spin waves in the Co$_{40}$Fe$_{40}$B$_{20}$(20nm)/Ni$_{80}$Fe$_{20}$(26nm) bilayer system have been determined using wave-vector resolved Brillouin light scattering experiments\,\cite{G89,SSOH15} carried out on a plain film. To evidence the nonreciprocal character of these dispersions, both Stokes and anti-Stokes peaks have been recorded for the two possible polarities of the applied magnetic field, in the MSSW configuration ($\mathbf{H}_0 \perp \mathbf{k}$).

Figure~\ref{Fig_BLS}(a-c) shows BLS spectra obtained with $\mu_0H_0\!=\!+30$~mT, at different values of the in-plane wave vector $k$. Four different peaks, two Stokes and two anti-Stokes, can be identified, which are related to the first (red) and second (blue) MSSW branches. These peaks have been fitted to Lorentzian lines in order to extract their central frequencies and thereby reconstruct the $f(k)$ curves. The obtained dispersion relations are found in very good agreement with theoretical predictions [Fig.~\ref{Fig_BLS}(d)]. As required to fulfill our ambition to build a magnonic diode, a well defined frequency plateau is present at about 12.5~GHz in the dispersion of MSSW mode 0, which corresponds to the anti-Stokes peak with $k$-independent position in Fig.~\ref{Fig_BLS}(a-c). Clear experimental evidence is then provided for the occurrence of nonreciprocal slow spin waves in our system.

We note that the anti-Stokes peak with highest frequency has a small amplitude, whatever $k$ [Fig.~\ref{Fig_BLS}(a-c)]. The peak even becomes undetectable for wave-vector values exceeding 17~rad/$\mu$m, hence the lack of some (blue) data points in Fig.~\ref{Fig_BLS}(d,e). This is attributed to the fact that, as verified in numerical simulations (not shown), MSSW mode 1 has very small amplitude in the upmost part of the bilayer film, whose magneto-optic contribution dominates the BLS signal. We also note that the frequency nonreciprocities of the two MSSW modes, $\Delta f_n(k)=f_n(-|k|)-f_n(+|k|)$ ($n\!=\!0,1$) show very specific behaviors [Fig.~\ref{Fig_BLS}(e)]. First, $\Delta f_0(k)$ and $\Delta f_1(k)$ are almost equal in absolute value and opposite in sign, which comes naturally in our analytical theory of the nonreciprocal hybridization (Sec.~\ref{Sec_Theory}). Second, these quantities do not vary monotonously as a function of $k$. Instead, they exhibit a local extremum followed by a change of sign at about 5~rad/$\mu$m.

Following the same line of thought as used to explain the nonintuitive localization of dipole-exchange MSSW modes\,\cite{K13,HBKL14}, this change of sign can be ascribed to a transition between a regime dominated by exchange across the film thickness, in the $k\!\rightarrow\!0$ limit, to a regime where in-plane dipole fields gain importance, at larger $k$. Upon hybridization, indeed, out-of-plane exchange interactions and dipole interactions produce additional torques which compete with each other and yield terms with opposite signs in the expression of $\Delta\Omega$ (Eq.~\ref{Eq_Delta_Omega}). To better see this, we may expand the elements of the dynamic matrix $\bar{\bar{C}}$ (Eqs.~\ref{Eq_Omega} and \ref{Eq_beta}) in Taylor series around $kl\!=\!0$. Keeping only terms up to second order in $kl$, we obtain
\begin{equation}\label{Eq_Delta_Omega_Taylor}
\Delta\Omega(\beta)=\frac{8kl}{\pi^3[\Omega_1^2(0)\!-\!\Omega_0^2(0)]} \left[\frac{\Lambda^2\pi^2}{l^2}\!- \frac{|k|l}{2}\!\left(1\!+\!3\,\frac{\Lambda^2\pi^2}{l^2}\right)\right]\!\beta,
\end{equation}
where we identify the first term between the square brackets, i.e. $\Lambda^2\pi^2/l^2$, as being of pure exchange origin and the second one, proportional to the $k$-dependent self-demagnetizing factor $P_{00}(k)\simeq \frac{|k|l}{2} $, as arising from dipole-dipole interactions. According to this expression, a change of sign of $\Delta\Omega$ is expected at the particular value $|k^*|=[\frac{l}{2}(3+\frac{l^2}{\Lambda^2\pi^2})]^{-1}$, which indeed amounts to a few radians per micrometer for magnetic materials with $\Lambda\sim 4$~nm and thickness $l$ in the $40\!-\!50$~nm range. Interestingly, such a change of sign has also been predicted for films where breaking of the top/bottom magnetic symmetry is not obtained through a blockwise variation of $M_{\text{S}}$, like here, but rather by a grading of saturation magnetization across the thickness\,\cite{GASG19}. In contrast, it has not been observed for bilayers where the two magnetic materials, being separated by a non-magnetic spacer, are only coupled through long range dipole interactions\,\cite{MGLA17,GSCO19}. This further demonstrates the essential role played by short range out-of-plane exchange-coupling in this phenomenon.

\begin{figure}
\includegraphics[width=8.5cm,trim=25 140 80 130,clip]{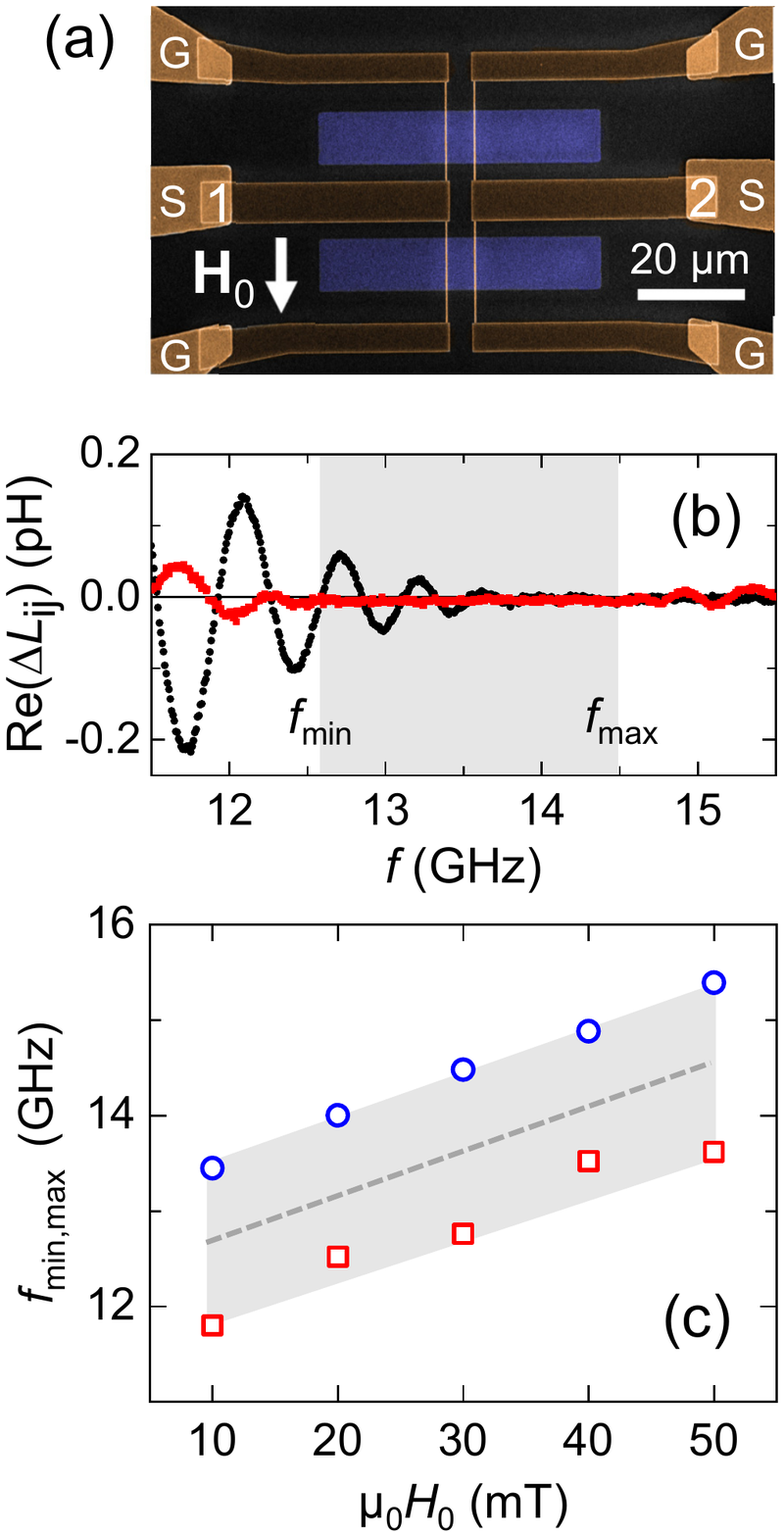}
\caption{Propagating spin-wave spectroscopy. (a) False color scanning electron micrograph of a PSWS device with a distance $D\!=\!5~\mu$m between the emitting and receiving antennas. (b) Real part of the spin-wave induced change in mutual inductance $\Delta L_{ij}$ as a function of the excitation frequency $f$ ($\mu_0H_0\!=\!+30$~mT), for spin waves propagating from antennas 1 to antennas 2 ($k\!>\!0$, red squares) and vice versa ($k\!<\!0$, black circles), in the device of panel (a). (c) Upper limit ($f_{\text{max}}$, blue circles) and lower limit ($f_{\text{min}}$, red squares) of the frequency gap for rightward propagating spin waves as a function of the applied magnetic field, as deduced from PSWS data such as shown in panel (b).} \label{Fig_PSWS_1}
\end{figure}

\subsection{Magnonic diode behavior}\label{Sec_Diode}
\subsubsection{Propagating spin-wave spectroscopy}\label{Sec_PSWS}
In order to clearly demonstrate the diode-like behavior of the studied bilayer system, propagating spin-wave spectroscopy (PSWS) experiments have been carried out on specially designed devices [Fig.~\ref{Fig_PSWS_1}(a)], each one containing a pair of $50~\mu$m-long, $10~\mu$m-wide bilayer waveguides and two pairs of single-wire antennas connected in parallel. This device layout, in which spin-waves are travelling simultaneously along two magnetic buses, was chosen to ensure a good symmetry match with Ground-Signal-Ground (GSG) microwave probes. Importantly, the 200~nm wide single-wire antennas used here can couple inductively to spin waves with a broad range of wave vectors, $0\!\leq\!|k|\!\leq\!k_{\text{max}}$, where $k_{\text{max}}\!\simeq\!12$~rad/$\mu$m. The distance between the emitting and receiving antennas ($D\!=2~\mu$m or $5~\mu$m) is adapted to the typical attenuation length expected for surface spin waves with such $k$ values in a Co$_{40}$Fe$_{40}$B$_{20}$(20nm)/Ni$_{80}$Fe$_{20}$(26nm) bilayer.

A typical PSWS experiment, as reported below, is performed in the following way. A vector network analyzer (VNA) is connected to the antennas through GSG probes to serve both as a generator and as a detector for determining the complex mutual inductance $L_{ij}$ of the two antennas pairs. Upon injection of a current with adequate frequency $f$ in the emitting antennas (index $j$), these couple inductively to the magnetization of the waveguides and spin waves are excited. If those waves travel far enough and reach the receiving antennas (index $i$) before being fully damped, a microwave magnetic flux with frequency $f$ is picked up. The ratio of the measured flux to the injected current defines the mutual inductance $L_{ij}$ of interest\,\cite{VB08}. In practice, in order to extract spin-wave related signals more accurately, relative measurements are systematically taken, wherein a background signal, which is acquired at much larger applied magnetic field so that no spin wave resonance occurs, is subtracted from the raw data.

Figure~\ref{Fig_PSWS_1}(b) shows the real part of the spin-wave induced change in mutual inductance $\Delta L_{ij}$ as a function of frequency for a device with a relatively large distance between the emitting and receiving antennas [$D\!=5\,\mu$m, see Fig.~\ref{Fig_PSWS_1}(a)], submitted to a transverse in-plane magnetic field $\mu_0H_0\!=\!+30$~mT. The two data sets presented correspond to opposite directions of spin-wave propagation. Comparing them, one immediately sees that the spin-wave signal at $12.5~\text{GHz}\!\leq\!f\!\leq\!14.5$~GHz is vanishingly small for $k\!>\!0$ (red symbols), meaning that no spin wave travel from the left antennas to the right antennas, whereas it is comparatively large for $k\!<\!0$ (black symbols) as spin waves do propagate effectively from the right antennas to the left ones. This nonreciprocal behavior is of course related to the presence of a plateau in the positive-$k$ part of the dispersion relation of MSSW mode 0 [Fig.~\ref{Fig_BLS}(d)], which has two main consequences. First, rightward propagating spin waves with $f\!\sim\!12.5$~GHz and $0\!\leq\!k\!\leq\!k_{\text{max}}$ are excited by the left antennas but, due to their very low group velocity, they die out under the effect of magnetic damping before reaching the receiving antennas. Second, due to the large extension of the plateau, the group velocity of MSSW mode 0 becomes sizable again only for $k$ values, which lie far beyond the $k$ range accessible with the used antennas, meaning that rightward propagating MSSW 0 spin waves with frequency well above 12.5~GHz are simply not produced. For $f\!\geq\!14.5$~GHz, MSSW mode 1 eventually gets excited so that a clear spin-wave signal is transmitted again for both directions of propagation. An effective forbidden gap with a width of about 2~GHz is thus formed for rightward propagating spin waves [shaded zone in Fig.~\ref{Fig_PSWS_1}(b)]. In view of the possible application of this phenomenon, it is worth mentioning that the gap can naturally be shifted up and down in frequency by adjusting the amplitude of the applied magnetic field. A tunability of the order of 50~MHz/mT could be measured experimentally over the 10-50~mT range [Fig.~\ref{Fig_PSWS_1}(c)].

\begin{figure}
\includegraphics[width=8.5cm,trim=25 160 25 220,clip]{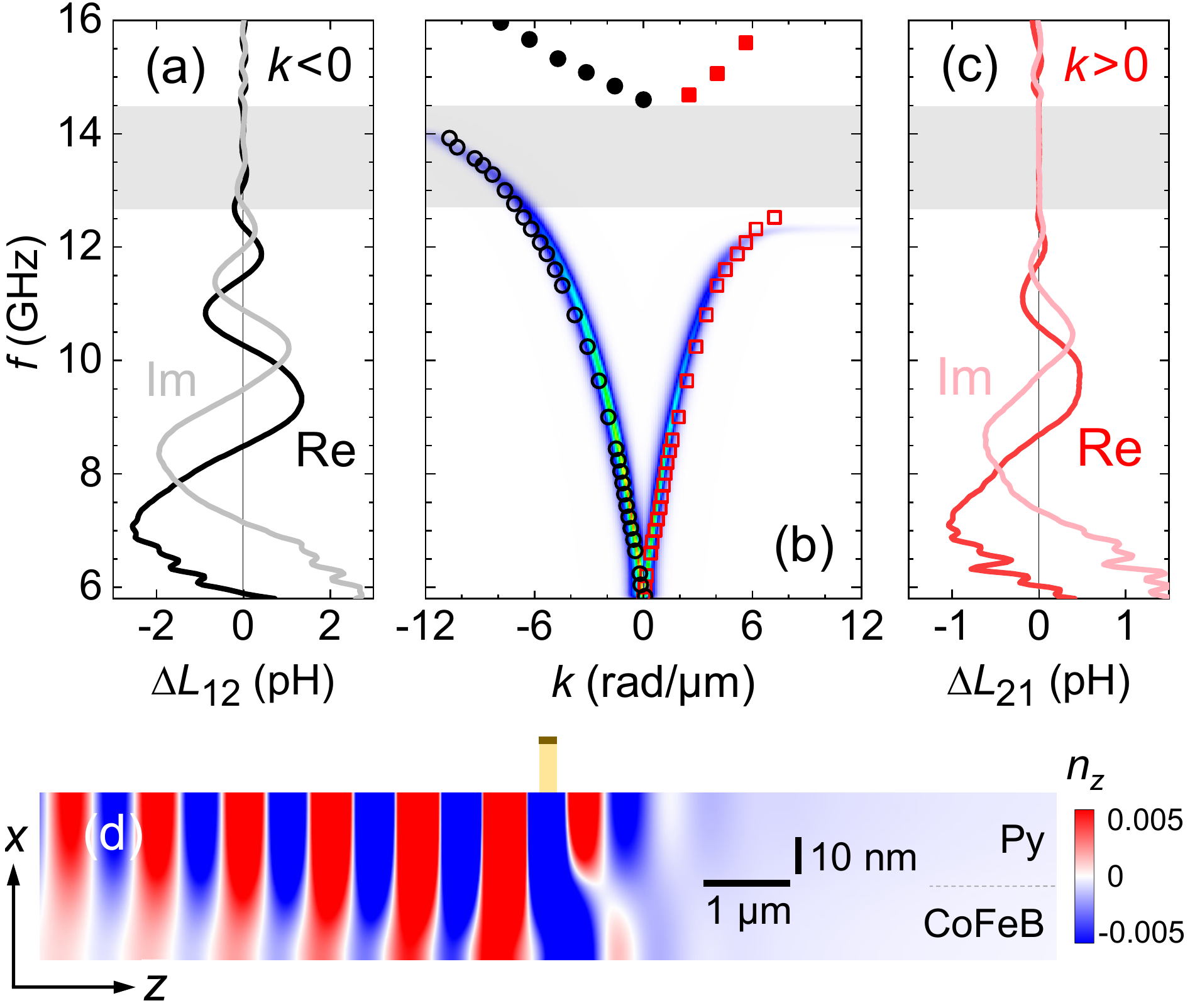}
\caption{Spin-wave propagation in a device with $D\!=\!2~\mu$m. (a,c) Real and imaginary parts of $\Delta L_{ij}$ as a function of frequency for $k\!<\!0$ (a) and $k\!>\!0$ (c). (b) Dispersion relations of the lowest (open symbols) and second lowest (solid symbols) MSSW branches deduced from the experimental spin-wave signals shown in (a) and (c), and spin-wave spectral weight (color map) as obtained from MuMax3 micromagnetic simulations. See text for details. (d) Simulated cross-sectional map of the in-plane component of the normalized dynamic magnetization, $n_z\!=\!m_z/M_{\text{S}}$, for an excitation frequency of 12.5~GHz. The lateral extension of the single-wire antenna is indicated by the brown horizontal bar and the insulating layer of silicon oxide is sketched as a yellow layer (not to scale vertically). The AC current assumed in the antenna is 3~mA, which yields an in-plane magnetic field of about 3~mT at the top surface of the bilayer, right beneath the source.} \label{Fig_PSWS_2}
\end{figure}

Because the wave vector $k$ is constrained to change according to the dispersion relation $k(f)$, the phase delay $kD$ acquired by spin waves after propagation over the distance $D$ varies continuously as the frequency $f$ is swept. This variation in phase delay translates into pronounced oscillations of the recorded spin-wave signal [Fig.~\ref{Fig_PSWS_1}(b)]. As we shall describe below, this provides us with a way to extract the wave vector value corresponding to each driving frequency $f$. For that, one needs to record both the real and imaginary parts of the spin-wave induced change in mutual inductance over a large range of frequencies [Fig.~\ref{Fig_PSWS_2}(a,c)] encompassing the ferromagnetic resonance frequency (FMR), $f_{\text{FMR}}=f_0(k\!=\!0)$, which corresponds to the onset of the oscillations\,\cite{Rem5}. From such data, the spin-wave wave vector can be determined as
\begin{equation}\label{Eq_SW_Phase}
k(f)=\pm\frac{\phi_{ij}(f)-\phi_0}{D}.
\end{equation}
In this expression, the $\pm$ sign accounts for the change of sign of $k$ upon reversing the direction of spin-wave propagation [$+$ corresponding to spin waves travelling from the left antennas ($j\!=\!1$) to the right ones ($i\!=\!2$)], $\phi_0\!=\!\frac{\pi}{2}$ is the reference phase at $f_{\text{FMR}}$, where $\Delta L_{ij}$ is purely imaginary (pure absorption), and, more important, $\phi_{ij}(f)\!=\arg[\Delta L_{ij}(f)]+2n\pi$ (with $n$ integer) is the spin-wave phase, which must be unwrapped in a continuous manner, starting from $f_{\text{FMR}}$.

The open symbols in Fig.~\ref{Fig_PSWS_2}(b) show the dispersion relation of MSSW mode 0 reconstructed by applying the above method (Eq.~\ref{Eq_SW_Phase}) to $\Delta L_{ij}$ data recorded between 5.8 and 14.5~GHz\,\cite{Rem6}. As expected, for $k\!>\!0$, the dispersion can only be followed up to approximately +7~rad/$\mu$m, which corresponds to the lower edge of the frequency plateau. In contrast, for $k\!<\!0$, it can be followed down to -12~rad/$\mu$m, thus confirming the ability of our PSWS device to probe the expected wave-vector range $[-k_{\text{max}},+k_{\text{max}}]$. As a support to our conclusions, Figure~\ref{Fig_PSWS_2}(b) also displays in the background the "weighted" dispersion relation computed for a CoFeB(20nm)/Py(26nm) bilayer with the MuMax3 software using space and time Fourier transforms of in-plane magnetization traces obtained under square pulse excitation of 10~ps duration. The simulations, which use periodic boundary conditions in both longitudinal ($z$) and transverse ($y$) in-plane directions and cell sizes $h_x\!=\!0.5$~nm, $h_y\!=\!10$~nm, $h_z\!=\!4$~nm, include a realistic spatial distribution for the excitation field produced by the antenna and account for magnetic losses through Gilbert damping factors of 0.008 and 0.012 for CoFeB and Py, respectively (see Sec.~\ref{Sec_muBLS}). A very good agreement between the two kinds of data may be observed, particularly regarding the asymmetric way in which the amplitude of the transmitted spin-wave vanishes upon increasing $f$. Figure~\ref{Fig_PSWS_2}(d) shows a cross-sectional map of the dynamic magnetization as generated under continuous-wave excitation at $f\!=\!12.5$~GHz, which illustrates how this nonreciprocity translates in real space: Once the excitation frequency enters the gap, the dynamic magnetization profile becomes essentially evanescent on the right side of the source as the far-field coupling of the antenna to the magnetic precession vanishes. Noticeably, micromagnetic simulations reveal a similar behavior in the "overshoot" regime, where the dispersion of MSSW mode 0 contains a region with backward character [Fig.~\ref{Fig_Principle}(d)]: The lower frequency limit of the gap then corresponds to the local maximum in the $\omega(k)$ curve of MSSW mode 0.

\begin{figure*}
\includegraphics[width=16cm,trim=50 90 20 90,clip]{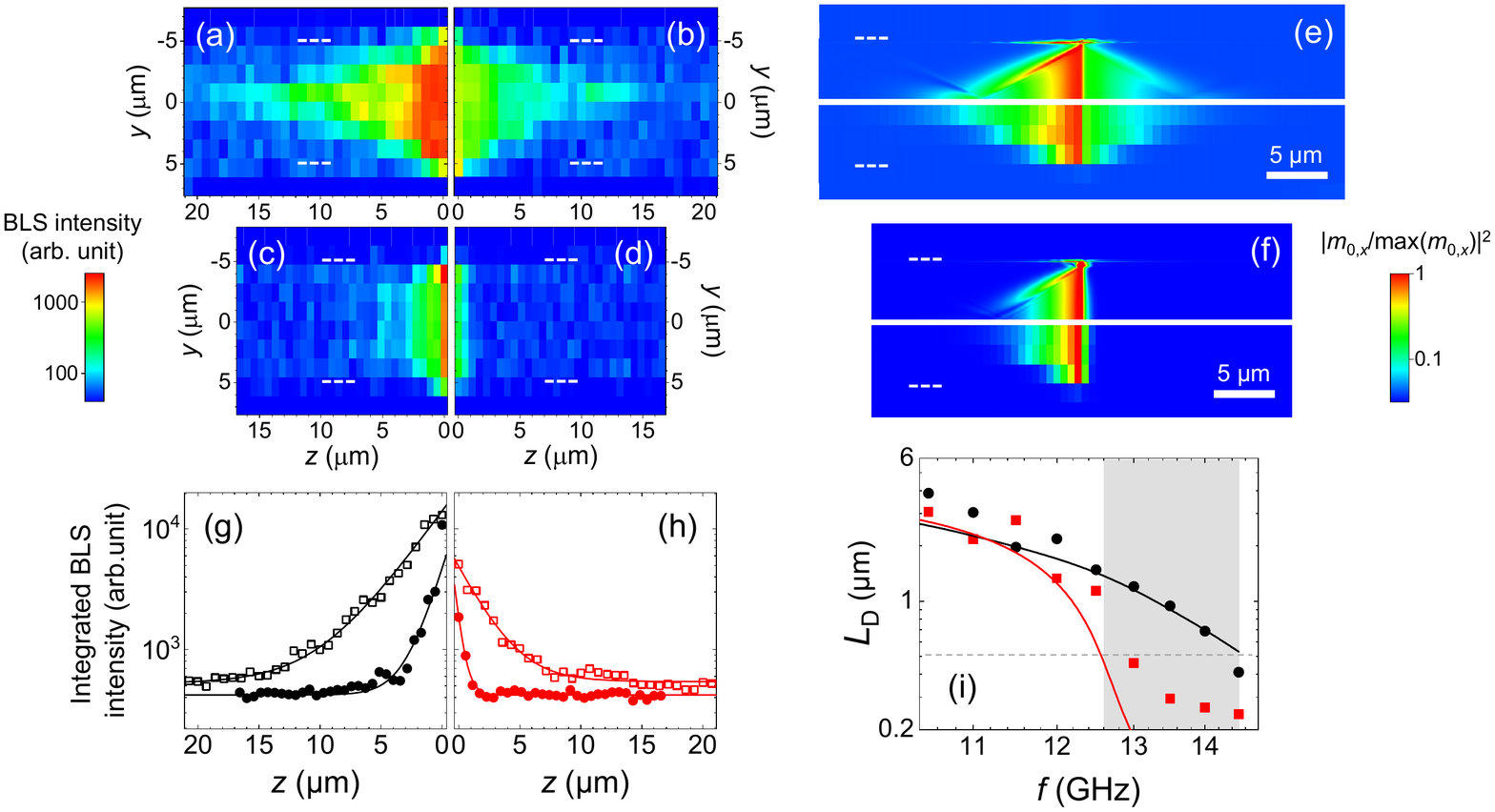}
\caption{(a-d) Experimental BLS intensity maps for excitation frequencies of 11~GHz (a,b) and 13~GHz (c,d) and applied magnetic fields $\mu_0H_0\!=\!-30$~mT (a,c) and $\mu_0H_0\!=\!+30$~mT (b,d) (logarithmic scale). The dashed lines indicate the position of the waveguide edges. (e,f) Computed BLS intensity maps for $f\!=\!11$~GHz (e) and $f\!=\!13$~GHz (f), $\mu_0H_0\!=\!+30$~mT. The current assumed in the antenna is 0.1~mA (linear regime). In each panel, the upper part shows raw data whereas the lower part shows pixelated data obtained by averaging raw data over rectangular areas. See text for further details. (g,h) $z$-profiles of the BLS intensity integrated over the width of the waveguide as deduced from the maps shown in panels (a-d). Open and solid symbols correspond to $f\!=\!11$~GHz and $f\!=\!13$~GHz, respectively. The lines are fits of the experimental data to the expression $I(z)=I_0\exp(-z/L_{\text{D}})+I_{\text{Noise}}$. (i) Variation of the decay length of the spin-wave intensity, $L_{\text{D}}$, with the excitation frequency $f$ for spin waves propagating to the right ($H_0\!>\!0$, red) and to the left ($H_0\!<\!0$, black). The symbols and the solid lines are data derived from experiments and simulations, respectively. The horizontal dashed line indicates the smallest spin-wave wavelength compatible with the chosen antenna design.} \label{Fig_muBLS}
\end{figure*}

\subsubsection{Micro-focussed Brillouin Light scattering}\label{Sec_muBLS}
To visualize directly the spatial decay of spin waves, we have performed micro-focussed BLS imaging\,\cite{SSOH15} on a similar device as used for PSWS [Fig.~\ref{Fig_PSWS_1}(a)]. In those experiments, the spin-wave intensity has been mapped next to an antenna while microwave power (-5~dBm) was continuously injected into it. In the DE configuration, switching the direction of the equilibrium magnetization is equivalent to reversing the wave vector $\mathbf{k}$\,\cite{HBKL14,C87}. Then, instead of looking on both sides of the source for imaging counterpropagating waves, we have concentrated ourselves on one side and recorded spin-wave intensity maps for the two polarities of the transversally applied magnetic field. Accordingly, the data obtained for $H_0\!<\!0$ are mirrored horizontally in Fig.~\ref{Fig_muBLS}(a,c). The benefit of this experimental strategy is that it allows us probing counterpropagating spin-waves in the very same optical conditions, thus avoiding artifacts related, for instance, to differences in the surface state of the waveguide.

To support these observations, MuMax3 simulations have also been performed for a finite, $50~\mu$m-long, $10~\mu$m-wide bilayer strip. The expected BLS signal has been calculated by assuming that it is mostly related to the out-of-plane component of the dynamic magnetization ($m_x$) at the top surface of the magnetic medium. As in the experiments, spin waves were excited by an alternating magnetic field with frequency $f$ produced by a 100~nm thick, 200~nm wide antenna, located 120~nm above the spin-wave conduit. For each magnetic cell (with size $h_x\!=\!4$~nm, $h_y\!=\!40$~nm, $h_z\!=\!8$~nm), the time dependence of $m_x$ was recorded over a full period $1/f$, in the steady excitation regime, and analyzed to extract its maximum value $m_{0,x}$. Finally, normalized intensity maps were constructed, which show $|m_{0,x}/\text{max}(m_{0,x})|^2$, either with the full resolution of the simulations [top panels in Fig.~\ref{Fig_muBLS}(e,f)] or with a degraded resolution mimicking that of micro-focussed BLS images [bottom panels in Fig.~\ref{Fig_muBLS}(e,f)]. Overall, a good agreement is obtained between experimental and computed images, assuming damping values of 0.008 and 0.012 for CoFeB and Py, respectively\,\cite{Rem7}. Quite naturally, upon pixelation, sharp features, like those related to localized edge modes, tend to be washed out. Yet, we note that for $f=11$~GHz, a long oblique contrast arising from the interference between the fundamental and higher-order width modes of the waveguide remain discernible [Fig.~\ref{Fig_muBLS}(a,e)].

Expectedly, for frequencies in the range 6-12~GHz (i.e. below the frequency plateau), significant spin-wave intensity is systematically detected up to distances of several micrometers from the antenna [Fig.~\ref{Fig_muBLS}(a,b,e)]. We note that although plateau-related spin-wave filtering is not yet active, a difference in intensity may be observed between the two directions of propagation. This is nothing but the usual amplitude nonreciprocity of magnetostatic surface waves, which follows from the fact that a transducer placed on one side of a waveguide couples differently to counterpropagating surface waves\,\cite{SSNH08}. When the frequency reaches 13~GHz (above the frequency plateau), on the other hand, the spin-wave intensity remains relatively large for one direction of propagation [Fig.~\ref{Fig_muBLS}(c,f)] but drops abruptly for the opposite one [Fig.~\ref{Fig_muBLS}(d,f)]. Based on the discussion above, we naturally attribute this fast drop in intensity to the phenomenon of spin-wave slow-down associated with the presence of a plateau in the positive-$k$ part of the dispersion relation of MSSW mode 0.

This phenomenon is best illustrated by extracting the spin-wave decay length $L_{\text{D}}$ from the BLS data. For this, one may simply average the spin-wave intensity over the width of the waveguide (i.e. along $y$) so as to mitigate finite-width effects\,\cite{Rem8}, plot the integrated intensity as a function of the space coordinate along the direction of propagation, $z$, and fit this dependence to an exponential decay of the form $I(z)=I_0\exp(-z/L_{\text{D}})+I_{\text{Noise}}$ [see Fig.~\ref{Fig_muBLS}(g,h)]. Figure~\ref{Fig_muBLS}(i) shows the variation of $L_{\text{D}}$ with $f$ obtained treating both experimental and numerical data in this manner. A clear difference in behavior may be observed depending on the sign of $k$. For spin waves travelling to the left ($k\!<\!0$, black circles and line), $L_{\text{D}}$ decreases steadily with increasing $f$. For spin waves travelling to the right ($k\!>\!0$, red squares and line), in contrast, this steady decay is interrupted by a sudden drop in $L_{\text{D}}$ as $f$ reaches the frequency of the plateau, $f_{p}$. Beyond $f_{p}$, $L_{\text{D}}$ becomes smaller than the minimum spin-wave wavelength $2\pi/k_{\text{max}}\simeq 0.5~\,\mu$m attainable with our single-wire antenna. This reveals the evanescent character of the magnetization dynamics induced when the frequency falls into the effective gap, also evidenced in the micromagnetic simulation of Fig.~\ref{Fig_PSWS_2}(d).

\section{Conclusion}\label{Sec_Conclusion}
In the present work, a new concept of spin-wave diode is proposed, which makes use of the particular dynamic dipolar interactions in transversally magnetized media. The device is made from a thin film consisting of two exchange-coupled layers with different saturation magnetization values. Our theoretical analysis reveals that, in such a film, chiral dipolar couplings develop, which results in nonreciprocal hybridization between close-lying spin-wave branches. Using this phenomenon, we engineer carefully the dispersion of surface waves so as to reduce the group velocity of waves travelling in a particular direction to a very low value (slow waves) while maintaining a large value for those propagating the other way. A comprehensive experimental picture of the diode functioning is obtained by combining propagating spin-wave spectroscopy with Brillouin light scattering, in both reciprocal-space thermal mode and real-space imaging mode: The magnetization dynamics excited by a source of finite size take the form of genuinely propagating waves in the forward direction of the diode and reduce to evanescent waves in the reverse one. By design, our spin-wave diode is quite versatile as its operational frequency window can be adjusted by tuning the amplitude of the applied magnetic field and its forward and reverse directions can be interchanged by switching the polarity of the field. Since our concept of spin-wave diode relies on fully built-in rather than transduction-related nonreciprocity, it could contribute decisively to the advance towards the all-magnon approach for computing\,\cite{C19}. As a concluding remark, we wish to point out that engineering of unidirectionally slow spin waves could also prove useful in more general situations where long interaction times are needed within a limited space, for instance, for promoting nonreciprocal nonlinear coupling in the channel of a magnonic transistor\,\cite{CSH14}.

\begin{acknowledgments}
This work was funded by the French National Research's Agency (ANR) through the Programme d'Investissement d'Avenir under contract ANR-11-LABX-0058\_NIE within the Investissement d'Avenir program ANR-10-IDEX-0002-02. The authors acknowledge the STnano cleanroom facility for technical support and the High Performance Computing center of the University of Strasbourg for access to computing resources, part of which were funded by the EquipEx Equip@Meso project of Programme Investissements d'Avenir.

\vspace{0.5cm}
M.Gr. and M.Ge. contributed equally to this work. Y.H. and D.S. performed the micromagnetic simulations. M.B. and D.L. developed the analytical model. M.H. grew the films. M.Gr. and D.L. fabricated the devices and carried out the inductive measurements. M.Ge. performed the BLS experiments. M.B., P.P. and Y.H. supervised the project. Y.H., M.Gr., M.Ge., M.B. and P.P. wrote the manuscript. All authors discussed the results.
\end{acknowledgments}

\appendix*
\section{}\label{Sec_Appendix}
In this appendix, we derive analytical expressions for the additional elements, which appear in the dynamical matrix for magnetostatic surface waves upon top-bottom disymmetrization of the magnetic film through blockwise variations of the saturation magnetization and exchange constant values (see Sec.~\ref{Sec_Theory}). In the linearized Landau-Lifshitz equation (Eq.~\ref{Eq_LL}), the perturbation thus introduced is described by the term
\begin{equation}
\gamma\mu_0\langle M_{\text{S}}\rangle\!\int_{\!-l/2}^{l/2}\! \bar{\bar{G}}_k(x\!-\!x')\,a(x')\,[\boldsymbol{\eta}(x')\times\hat{\mathbf{y}}]\,dx'.
\end{equation}
Therefore, the corrections to be added to the $\bar{\bar{C}}$ matrix are four $2\!\times\!2$ blocks, each of which has the form
\begin{equation}
\bar{\bar{A}}_{nm}=
\begin{pmatrix}
-A_{nm}^{zx}&-A_{nm}^{zz}\\
A_{nm}^{xx}&A_{nm}^{xz}\\
\end{pmatrix},
\end{equation}
with
\begin{equation}\label{Eq_Anm}
A_{nm}^{ij}=\int_{\!-l/2}^{l/2} dx \int_{\!-l/2}^{l/2} dx' S_n(x)\,G_k^{ij}(x-x')\,a(x')S_m(x').
\end{equation}
Here, indices $n,m\!=\!0,1$ refer to the uniform and antisymmetric basis functions introduced in Sec.~\ref{Sec_Theory} and the $G_k^{ij}$'s ($i,j\!=\!x,z$) are the four components of the tensorial magnetostatic Green's function, which read\,\cite{KS86}
\begin{eqnarray}\label{Eq_Green's_function}
G_k^{zz}(s)&=&-\frac{|k|}{2}e^{-|ks|}\nonumber\\
G_k^{xx}(s)&=&-\delta(s)-G_k^{zz}(s)\\
G_k^{xz}(s)&=&\,G_k^{zx}(s)=-i\,\text{sgn}(ks)\,G_k^{zz}(s), \nonumber
\end{eqnarray}
with $\delta(s)$ the Dirac's function. Since the functions $a(x)$, $G_k^{xz}(s)$, $G_k^{zx}(s)$, and $S_1(x)$ are odd and $G_k^{xx}(s)$, $G_k^{zz}(s)$, and $S_0(x)$ are even, half of these matrix elements are strictly nil: $A_{nn}^{ii}\!=\!A_{n\neq m}^{i\neq j}\!=\!0$. Moreover, from Eqs.~\ref{Eq_Anm} and \ref{Eq_Green's_function}, it is easy to see that one necessarily has $A_{nn}^{ij}\!=\!A_{nn}^{ji}$. Then, overall, only four independent quantities must be calculated: $A_{00}^{xz}$, $A_{11}^{xz}$, $A_{01}^{xx}$, and $A_{01}^{zz}$.

\begin{widetext}
\noindent Let us start with $A_{00}^{xz}$ and $A_{11}^{xz}$, which both involve $G_k^{xz}$. According to Eqs.~\ref{Eq_Anm} and \ref{Eq_Green's_function}, $A_{00}^{xz}$ writes
\begin{eqnarray}\label{Eq_A00xz_1}
A_{00}^{xz}
&=&
\int_{\!-l/2}^{l/2} dx \int_{\!-l/2}^{l/2} dx' S_0(x)\,G_k^{xz}(x-x')\,a(x')S_0(x')\\
&=&
i\frac{\beta|k|}{2l}\int_{\!-l/2}^{l/2} dx \int_{\!-l/2}^{l/2} dx' \text{sgn}[k(x-x')]e^{-|k(x-x')|}\text{sgn}(x').\nonumber
\end{eqnarray}
After some long but straightforward algebra for dealing with the absolute value and sign functions, we obtain
\begin{equation}\label{Eq_A00xz_2}
A_{00}^{xz} = i P_{00}' = i\frac{\beta}{kl}\left[\sinh(|k|l)-2\sinh(|k|l/2) \right].
\end{equation}
Similarly, for $A_{11}^{xz}$, we have
\begin{eqnarray}\label{Eq_A11xz_1}
A_{11}^{xz}
&=&
\int_{\!-l/2}^{l/2} dx \int_{\!-l/2}^{l/2} dx' S_1(x)\,G_k^{xz}(x-x')\,a(x')S_1(x')\\
&=&
i\frac{\beta|k|}{l}\int_{\!-l/2}^{l/2} dx \sin\left(\frac{\pi x}{l}\right) \int_{\!-l/2}^{l/2} dx' \text{sgn}[k(x-x')]e^{-|k(x-x')|}\text{sgn}(x') \sin\left(\frac{\pi x'}{l}\right), \nonumber
\end{eqnarray}
from which it comes
\begin{equation}\label{Eq_A11xz_2}
A_{11}^{xz} = i P_{11}' = i\frac{2\beta kl}{\pi^2+k^2l^2}\left(\pi-|k|l\,e^{-|k|l/2} \right).
\end{equation}

As made explicite in Eq.~\ref{Eq_Green's_function}, the $xx$ component of the tensorial Green's function is composed of two terms, a local term ($\delta$ function) corresponding to the usual, $k$-independent out-of-plane demagnetizing field and a nonlocal, $k$-dependent correction involving $G_k^{zz}$. Accordingly, the matrix element $A_{01}^{xx}$ may be decomposed as
\begin{equation}\label{Eq_A01xx}
A_{01}^{xx}\!=\!-I'+Q',
\end{equation}
with
\begin{eqnarray}\label{Eq_Ip_1}
I'
&=&
\int_{\!-l/2}^{l/2} dx \int_{\!-l/2}^{l/2} dx' S_0(x)\,\delta(x-x')\,a(x')S_1(x')\\
&=&
\frac{\beta\sqrt{2}}{l}\int_{\!-l/2}^{l/2} dx \int_{\!-l/2}^{l/2} dx' \,\delta(x-x')\,\text{sgn}(x')\sin\!\left(\frac{\pi x'}{l}\right) \nonumber
\end{eqnarray}
and
\begin{eqnarray}\label{Eq_Qp_1}
Q'
&=&
\int_{\!-l/2}^{l/2} dx \int_{\!-l/2}^{l/2} dx' S_0(x)\,G_k^{zz}(x-x')\,a(x')S_1(x')\\
&=&
\frac{\beta|k|}{\sqrt{2}\,l}\int_{\!-l/2}^{l/2} dx \int_{\!-l/2}^{l/2} dx' \,e^{-|k(x-x')|}\,\text{sgn}(x')\sin\!\left(\frac{\pi x'}{l}\right). \nonumber
\end{eqnarray}
As before, the double integrations in Eqs.~\ref{Eq_Ip_1} and \ref{Eq_Qp_1} require only basic algebra and we obtain
\begin{equation}\label{Eq_Ip_2}
I' = \frac{2\sqrt{2}\beta}{\pi}
\hspace{1cm}\text{and}\hspace{1cm}
Q' = \frac{2\sqrt{2}\beta}{\pi^2+k^2l^2}\left[  \pi\!\left( 1\!-\!e^{-|k|l\!/2} \right)+ k^2l^2\!\left( \frac{1}{\pi}-\frac{1\!-\!e^{-|k|l}}{2|k|l} \right) \right]
\end{equation}
\noindent Finally, for $A_{01}^{zz}$, we simply have
\begin{equation}\label{Eq_A01zz}
A_{01}^{zz} =
\int_{\!-l/2}^{l/2} dx \int_{\!-l/2}^{l/2} dx' S_0(x)\,G_k^{zz}(x-x')\,a(x')S_1(x') = Q'.
\end{equation}
\end{widetext}


\begin{thebibliography}{0}
\bibitem{Rem1}
A similar behavior can be observed for magnetic films containing a smooth gradient of $M_{\text{S}}$ through the thickness, see Ref.~\onlinecite{GASG19}. These are however much more difficult to fabricate experimentally and to model analytically.
\bibitem{Rem2}
Throughout this entire paper, we assume $\gamma/2\pi=29.2$~GHz/T, which is the value determined from ferromagnetic resonance experiments performed on single layers of Co$_{40}$Fe$_{40}$B$_{20}$ and Permalloy.
\bibitem{Rem3}
Exchange modes are defined here as eigenmodes of the exchange operator, see Ref.~\onlinecite{KS86}.
\bibitem{Rem5}
The short-period small-amplitude oscillations of the mutual inductance observed at frequencies between 6 and 7~GHz are related to reflections of the spin waves at the extremities of the magnetic waveguides.
\bibitem{Rem7}
These values are significantly larger than those determined for CoFeB (0.004) and Py (0.006) from usual FMR experiments performed on as-deposited single layers of those materials. Yet, these are the values that must be assumed to reach a good agreement when trying to reproduce, in numerical simulations, the spatial decay of spin waves observed experimentally through microfocused Brillouin Light Scattering. Such an increase in damping could be related to the nanofabrication process.
\bibitem{Rem8}
Interferences between width modes of the waveguide with even symmetries, in particular the fundamental mode ($n=0$) and the third width mode ($n=2$), are washed out when averaging the spin-wave intensity over the width of the magnetic strip, so that the decay length extracted here are not influenced by such effects.
\end{thebibliography}
\end{document}